# Electronic Structure and Transport Properties of Cu-deficient Kuramite $Cu_{3-x}SnS_4$

Yosuke Goto*, Yuki Sakai, Yoichi Kamihara, and Masanori Matoba

*Department of Applied Physics and Physico-Informatics, Faculty of Science and Technology, Keio University, 3-14-1 Hiyoshi, Yokohama 223-8522, Japan*

E-mail: ygoto@z8.keio.jp

**Abstract** Electrical and thermal transport properties of Cu-deficient kuramite $Cu_{3-x}SnS_4$ (CTS) was examined as a possible earth-abundant thermoelectric material. Crystallographic structure of CTS was characterized by partial disorder between Cu and Sn. In contrast to semiconducting electrical transport of related compounds, such as $Cu_2ZnSnS_4$ and $Cu_3SbS_4$, metallic conduction with an electrical resistivity of 0.4 mΩcm and a carrier concentration of $3 \times 10^{21}$ cm$^{-3}$ was observed at 300 K. Lattice thermal conductivity was calculated at ~2 Wm$^{-1}$K$^{-1}$, which was probably reduced by Cu-deficiency and/or partial cation disorder. Density functional theory calculation indicates valence band was composed of hybridization between Cu $3d$ orbitals and S $3p$ orbitals.

## 1. Introduction

Since Thomas Seebeck discovered that a temperature gradient applied to a bimetallic loop deflects a magnetic needle,[1] much attention has been paid to this subject. Thermoelectric (TE) devices directly convert heat to electricity or pump heat using electricity, without any moving parts or gas/liquid fluids.[2] Currently, the applications of TE devices are limited to specialized fields such as deep-space power generation. Therefore, further investigations are required to improve TE efficiency for more widespread use of thermoelectrics. The efficiency of TE devices is determined by the Carnot efficiency and the dimensionless figure of merit of materials, $ZT = S^2 T \rho^{-1} \kappa^{-1}$, where $T$ is the absolute temperature, $S$ is the Seebeck coefficient, $\rho$ is the electrical resistivity, and $\kappa$ is the thermal conductivity.[3] Attempts to enhance the $ZT$ are often impeded by the fact that these transport properties strongly interact with one another.[4,5] However, researchers have demonstrated TE materials with $ZT > 1$ through various strategies such as increasing the power factor ($S^2 \rho^{-1}$) by the optimization of carrier concentrations and/or the electronic structure[6,7] and reducing $\kappa$ via the manipulation of crystallographic structure[8–10] or nanostructure engineering.[11,12]

Crystallographic structure of quaternary stannite/kesterite-type compounds $Cu_2$–II–IV–$Ch_4$, where II is a transition-metal element with the 3$d$ shell more than half-filled, IV is a group 14 element, and $Ch$ is a chalcogen, could be derived from the zincblende-type structure through symmetry decay, as shown in Figure 1.[13–16] Numerous elemental combinations of these compounds have generated interest because of their functionality as thin-film solar cell absorbers[17] and TE devices.[18–25] The intrinsic $p$-type conduction and relatively low $\kappa$ of these compounds result in fairly good TE properties, with $ZT$ values as high as 0.95 at 850 K for $Cu_2ZnIn_{0.1}Sn_{0.9}Se_4$.[18] Computational simulations have predicted that the intrinsic $p$-type conduction of these compounds originates from Cu vacancies; *i.e.*, Cu has a much lower defect formation energy than the other constituent elements.[26] The $\kappa$ of these quaternary compounds is ~4 $Wm^{-1}K^{-1}$, which is much lower than that of binary zincblende[27] because of mass contrast. Furthermore, Neutron diffraction and synchrotron X-ray diffraction (SXRD) revealed that the existence of anti-site disorder of Cu and Zn in $Cu_2ZnSnS_4$.[28,29] This anti-site disorder might reduce the $\kappa$ due to point-defect scattering, however, the similar number of electrons for transition metals such as Cu and Zn complicates the investigation of cation contribution via conventional X-ray diffraction (XRD). We have demonstrated that the cation contribution in kuramite $Cu_3SnS_4$ (CTS)[30,31] was easily detected using conventional XRD because the number of electrons in Cu and Sn were substantially different.[32] Indeed, crystal structure of CTS was characterized by partial disorder between Cu and Sn. Although thin films[33,34] and nanostructured[35,36] CTS have been prepared and magnetization[37,38] and spectroscopic[39,40] measurements have been conducted on them, many of their electrical and thermal transport properties are not established, especially those of above 300 K. In this study, the electrical and thermal transport properties, optical properties, and electronic structure of $Cu_{3-x}SnS_4$ for nominal $x$ ($x_{nom}$) of 0.3 were investigated.

## 2. Experimental methods

Polycrystalline $Cu_{3-x}SnS_4$ for $x_{nom}$ of 0.3 was synthesized by solid-state reaction in a sealed silica tube using CuS, $Cu_2S$, and $SnS_2$ as starting materials. In this article, $x_{nom}$ denote the stoichiometry of these starting materials. To obtain CuS, $Cu_2S$, and $SnS_2$, stoichiometric amounts of Cu (Kojundo Chemical, 99.9%), Sn (Kojundo Chemical, 99.99%), and S (Sigma-Aldrich, 99.98%) were heated at 700 °C for 10 h in an evacuated silica tube. A mixture of CuS, $Cu_2S$, and $SnS_2$ powders were pressed and heated in a sealed silica tube at 650 °C for 40 h. After the heat treatment, the sample was quenched in iced water. To obtain dense sample, the sample was consolidated at 1 GPa and 300 °C for 1 h by cubic anvil press apparatus (Try-Engineering) using pyrophyllite cell. The relative density of sample was calculated at more than 95%. We note the stoichiometric $Cu_3SnS_4$ ($x_{nom} = 0$) was not synthesized by this method; $x_{nom} = 0$ sample contains $Cu_{2-\delta}S$ impurity, indicating $x_{nom} = 0$ was highly Cu-deficient.

The purity of the sample was examined by XRD (Rigaku Rint 2500) using Cu Kα radiation. Reference-grade Si powder (NIST SRM 640d) was utilized as an external reference. The Rietveld refinement was performed using the RIETAN-FP code.[41] The crystal structure was drawn using the VESTA software.[42] The surface of samples was examined using scanning electron microscope (SEM; FEI Inspect S50).

Hall coefficient at 300 K was measured using the five-probe geometry under magnetic fields ($H$) up to $\mu_0 H = \pm 1$ Tesla. The thickness of the sample was sharpened to ~100 μm for the Hall measurement. The measurements of the $\rho$, $S$, and $\kappa$ were conducted at temperatures up to 623 K using a lamp heating unit (Ulvac, MILA-5000). The $\rho$ was measured by the dc four-probe technique using Pt wires attached by silver paste (DuPont 6838); the measurements were performed under a nitrogen atmosphere. The $S$ was obtained from the slopes of the plots of Seebeck voltages vs. temperature differences ($\Delta T$) measured by Pt–Pt/Rh 13% thermocouples. The $\kappa$ was obtained from the slopes of the plots of heat flux density vs. $\Delta T$ using a strain gauge as a heater and was measured under pressures less than $10^{-3}$ Pa. The heat loss by radiation through the sample[43] was subtracted under the assumption that emissivity is independent of temperature and wavelength during the measurements of $\kappa$. The emissivity of 0.8 was employed on the basis of reflectivity (*R*) measurements, which were performed at room temperature using a spectrometer equipped with an integrating sphere (Hitachi High-tech, U-4100). A photomultiplier tube and PbS were utilized as reflectance detectors in the wavelength ranges of 400–800 and 930–2600 nm, respectively. $Al_2O_3$ powders were used as standard reference materials. Absorption coefficient ($\alpha$) was converted from *R* using the Kubelka–Munk equation,[44] $(1-R)^2/2R = \alpha/s$, where *s* denotes the scattering factor.

The core-level and valence-band structures were measured by X-ray photoemission spectroscopy (XPS; JEOL JPS-9010-TR) on an instrument equipped with a Mg K$\alpha$ radiation. The sample surface was scraped using $Ar^+$-ion etching (0.5 kV, 90°, 60 s) in the preparation chamber before XPS measurements. The scale of the binding energy was calibrated on the basis of the 4*f* peaks of Au.

Theoretical band structure was computationally simulated using the plane-wave projector augmented-wave (PAW)[45,46] method implemented in the Vienna *ab initio* Simulation Package (VASP) code.[47,48] The exchange-correlation potential was approximated by the hybrid functionals using the Heyd, Scuseria, and Ernzerhof (HSE06) method.[49] The Brillouin zone was sampled by a $6 \times 6 \times 3$ Monkhorst–Pack grid,[50] and a cutoff of 450 eV was chosen for the plane-wave basis set. Hellmann–Feynman forces were reduced until 0.5 eV·$nm^{-1}$.

## 3. Results and discussion

### 3.1 Results

Figure 2 shows the XRD patterns and the corresponding Rietveld refinement of $Cu_{3-x}SnS_4$ for $x_{nom} = 0.3$. Almost all of diffraction peaks were assigned to those of tetragonal phase, except for several weak peaks due to $Cu_{2-\delta}S$ (2.0 wt.%) and SnS (1.5 wt.%), indicating that the CTS is the dominant phase in the sample. Refined structural parameters are summarized in Table 1. The 2*a* site was fully occupied by Cu, whereas the 2*b* and 4*d* sites were occupied by both Cu and Sn, as shown in Fig. 1. SEM images show that samples contain the ~1 μm of pores, as shown in the inset of Fig. 2.

Figure 3 shows electrical and thermal transport properties as a function of *T*. As shown in Figure 3(a), the $\rho$ was 0.4 mΩcm at 300 K and it increased with increasing *T*. This degenerate conduction is in contrast to semiconducting transport properties of related compounds such as $Cu_2ZnSnS_4$ (CZTS) and $Cu_3SbS_4$ (CAS). Indeed, the $\rho$ of CZTS and CAS at 300 K was reported as ~5 Ωcm and ~50 mΩcm, respectively.[20,51] The polarity of Hall measurement confirms that the dominant carrier of CTS is a conducting hole. The Hall carrier concentration ($n_H$) of $2.9(3) \times 10^{21}$ $cm^{-3}$ and Hall mobility of 5.3 $cm^2V^{-1}s^{-1}$ were observed at 300 K. Figure 3(b) shows the *S* as a function of *T*. The positive value of *S* indicates the samples are *p*-type conductor, consistent with the results of Hall measurement. The *S* generally increases with *T* and it reaches ~50 μVK$^{-1}$ at 623 K.

Figure 3(c) shows the $\kappa$ as a function of *T*. The $\kappa$ was 4.5(3) Wm$^{-1}$K$^{-1}$ at 300 K and it was almost independent of *T*. In a simple kinetic picture, the $\kappa$ is approximately expressed by the sum of electrical thermal conductivity ($\kappa_{el}$) and lattice thermal conductivity ($\kappa_l$). The $\kappa_{el}$ is described by Wiedemann-Franz law, $\kappa_{el} = L\rho^{-1}T$, where *L* is the Lorenz number. The $\kappa_{el}$ was calculated at ~2 Wm$^{-1}$K$^{-1}$, which is around 50% of the $\kappa$, on the basis of the constant value of *L*, 2.45 × 10$^{-8}$ V$^2$K$^{-2}$. The $\kappa_l$ of ~2.6 Wm$^{-1}$K$^{-1}$ at 300 K was lower than other related compounds.[20,51] This relatively low $\kappa$ was probably attributed to phonon scattering due to Cu-deficiency and/or partial cation disorder. Furthermore, the $\kappa_l$ of other copper-based chalcogenide is almost independent of Cu off-stoichiometry when the amount of Cu-deficiency is small,[52–54] suggesting that the reduced $\kappa_l$ of CTS arises from partial cation disorder, although such discussion requires single phase samples with high sintered density.

Figure 4 shows the XPS spectrum of CTS in the binding-energy range of 920 to 950 eV. The binding energy of Cu $2p_{3/2}$ (~933 eV) is similar to those of Cu metal and $Cu_2O$.[55,56] However, it distinctly differs from those of CuO,[56] suggesting that the Cu–S chemical bond in CTS is likely to be covalent or monovalent.

Figure 5 shows the valence-band XPS spectrum of CTS in the binding energy range of 0.4 to 10 eV, along with the calculated density of states (DOS) for stoichiometric $Cu_3SnS_4$. The XPS spectrum exhibits three distinct bands with peaks at approximately –2.7 eV, –4.1 eV, and –8.0 eV; these peaks are indexed as A–C. The theoretical DOS is consistent with valence-band structures measured by XPS. The bands A and B peak at approximately –2.7 eV and –4.1 eV and consist of Cu 3*d* and S 3*p* orbitals, respectively. The band C, which peaks at approximately –8.0 eV, is assigned to the hybridized orbitals of Sn 5*s* and S 3*p*. Although the ratio of the theoretical photoemission cross sections of the Cu 3*d*, Sn 4*s*, and S 3*p* states for Mg K$\alpha$ radiation ($h\nu$ = 1253.6 eV) is approximately 1.0: 0.1: 0.1,[57] an energy difference between Cu–S hybridized orbitals and Sn–S hybridized orbitals allows us to assign A–C structures in the XPS spectrum without correcting the cross section.

3.2 Discussion

The *ZT* value of CTS was less than 0.1 at 300–623 K. This value is quite small compared to the state of the art TE materials, however, we expect CTS could be a possible thermoelectric materials through optimization of chemical composition. For example, the $n_H$ of CZTS[20] and CAS[51,58] was reported as ~$10^{18}$ cm$^{-3}$, suggesting the $n_H$ of CZTS–CTS and CTAS–CTS solid solution could be controlled in the range of $10^{18}$ to $10^{21}$ cm$^{-3}$, which is suitable for thermoelectric materials because typical thermoelectric materials have optimized at the $n_H$ of $10^{19}$–$10^{20}$ cm$^{-3}$.

Figure 6(a) shows the $R$ spectra at room temperature. An increase in the $R$ below 0.8 eV indicates metallic reflection.[59] As shown in Figure 6(b), absorption edges were also observed at approximately 0.6 and 1.2 eV. The latter one might be attributed to transition from valence band to conduction band on the basis of calculated DOS (Fig. 5) because it is known that the calculation using HSE hybrid functional could predict the band gap of related compounds.[60] Notably, the band gaps of thin-film or nanostructured CTS have been reported to be 1.4–1.5 eV.[33–36]. We deduce the former one is assigned to transition due to defect states.[26] The optical properties of metals or heavily doped semiconductors are usually modeled using the Drude model; however, in this case, the absorption edges strongly overlap with the metallic reflection. Spectroscopic ellipsometry is expected to be helpful for further elucidation of the optical properties of CTS.[61]

## 4. Conclusions

Polycrystalline CTS was synthesized to demonstrate the electrical and thermal transport properties. Metallic conduction with $\rho$ of 0.4 mWcm and $n_H$ of 3 × $10^{21}$ cm$^{-3}$ at 300 K was in contrast to semiconducting transport of other stannite/kesterite-type compounds. Lattice thermal conductivity was probably reduced by Cu-deficiency and/or partial cation disorder. Valence band was composed of Cu 3$d$ and S 3$p$ hybridization.

### Acknowledgments

We thank Dr. Ikuya Yamada of Osaka Prefecture University for using high-pressure apparatus. This work was partially supported by the research grants from Keio University, the Keio Leading-edge Laboratory of Science and Technology (KLL), Japan Society for Promotion of Science (JSPS) KAKENHI Grant Numbers: 26400337, and Hitachi Metals Materials Science Foundation (HMMSF).

**References**

1) T. J. Seebeck, *Magnetische Porarisation der Metalle und Erze durch Temperature-Differenz* (Abh. Akad. Wiss. Berlin, 1822).
2) L. E. Bell, Science **321**, 1457 (2008).
3) A. F. Ioffe, *Semiconductor Thermoelements and Thermoelectric Cooling* (Infosearch Limited, London, 1957).
4) D. M. Rowe, *CRC Handbook of Thermoelectrics* (Chemical Rubber, Boca Raton FL, 1995).
5) G. D. Mahan, J. O. Sofo, Proc. Natl. Acad. Sci. USA **93**, 7436 (1996).
6) G. J. Snyder, E. S. Toberer, Nat. Mater. **7**, 105 (2008).
7) Y. Pei, X. Shi, A. LaLonde, H. Wang, L. Chen, G. J. Snyder, Nature **473**, 66 (2011).
8) B. C. Sales, D. Mandrus, R. K. Williams, Science **272**, 1325 (1996).
9) V. L. Kuznetsov, L. A. Kuznetsova, A. E. Kaliazin, D. M. Rowe, J. Appl. Phys. **87**, 7871 (2000).
10) E. S. Toberer, A. F. May, G. J. Snyder, Chem. Mater. **22**, 624 (2010).
11) L. D. Hicks, M. S. Dresselhaus, Phys. Rev. B **47**, 12727 (1993).
12) M. G. Kanatzidis, Chem. Mater. **22**, 648 (2010).
13) L. O. Brockway, Z. Kristallogr. **89**, 434 (1934).
14) C. H. L. Goodman, J. Phys. Chem. Solids **6**, 305 (1958).
15) B. R. Pamplin, J. Phys. Chem. Solds **25**, 675 (1964).
16) W. Schäfer, R. Nitsche, Mat. Res. Bull. **9**, 645 (1974).
17) M. T. Winkler, W. Wang, O. Gunawan, H. J. Hovel, T. K. Todorov, D. B. Mitzi, Energy Environ. Sci. **7**, 1029 (2014).
18) X. Y. Shi, F. Q. Huang, M. L. Liu, L. D. Chen, Appl. Phys. Lett. **94**, 122103 (2009).
19) M. L. Liu, I. W. Chen, F. Q. Huang, L. D. Chen, Adv. Mater. **21**, 3808 (2009).
20) M. L. Liu, F. Q. Huang, L. D. Chen, I. W. Chen, Appl. Phys. Lett. **94**, 202103 (2009).
21) C. Yang, F. Huang, L. Wu, K. Xu, J. Phys. D Appl. Phys. **44**, 295404 (2011).
22) M. Ibáñez, D. Cadavid, R. Zamani, N. García-castello, V. Izquierdo-roca, W. Li, A. Fairbrother, J. D. Prades, A. Shavel, J. Arbiol, A. Pe, J. R. Morante, A. Cabot, Chem. Mater. **24**, 562 (2012).
23) M. Ibáñez, R. Zamani, A. LaLonde, D. Cadavid, W. Li, A. Shavel, J. Arbiol, J. R. Morante, S. Gorsse, G. J. Snyder, A. Cabot, J. Am. Chem. Soc. **134**, 4060 (2012).
24) W. G. Zeier, A. Lalonde, Z. M. Gibbs, C. P. Heinrich, M. Panthöfer, G. J. Snyder, W. Tremel, J. Am. Chem. Soc. **134**, 7147 (2012).

25) W. G. Zeier, Y. Pei, G. Pomrehn, T. Day, N. Heinz, C. P. Heinrich, G. J. Snyder, W. Tremel, J. Am. Chem. Soc. **135**, 726 (2013).
26) S. Chen, J. H. Yang, X. G. Gong, A. Walsh, S. H. Wei, Phys. Rev. B **81**, 245204 (2010).
27) G. A. Slack, Phys. Rev. B **6**, 3791 (1972).
28) S. Schorr, H. Hoebler, M. Tovar, Eur. J. Miner. **19**, 65 (2007).
29) T. Washio, H. Nozaki, T. Fukano, T. Motohiro, K. Jimbo, J Appl. Phys. **110**, 074511 (2011).
30) V. A. Kovalenker, T. L. Evstigneeva, N. V. Troneva, L. N. Vyal'sov, Zap. Vses. Miner. Obsh. **108**, 564 (1979).
31) M. Fleisher, L. Cabri, G. Chao, Am. Mineral. **65**, 1065 (1980).
32) Y. Goto, F. Naito, R. Sato, K. Yoshiyasu, T. Itoh, Y. Kamihara, M. Matoba, Inorg. Chem. **52**, 9861 (2013).
33) P. A. Fernandes, P. M. P. Salomé, A. F. Da Cunha, J. Phys. D: Appl. Phys. **46**, 215403 (2010).
34) Z. Su, K. Sun, Z. Han, F. Liu, Y. Lai, J. Li, and Y. Liu, J. Mater. Chem. **22**, 16346 (2012).
35) Y. Xiong, Y. Xie, G. Du, and H. Su, Inorg. Chem. **41**, 2953 (2002).
36) X. Chen, X. Wang, C. An, J. Liu, and Y. Qian, J. Cryst. Growth **256**, 368 (2003).
37) F. Benedetto, T. Evstigneeva, M. Borgheresi, A. Caneschi, and M. Romanelli, Phys. Chemi. Miner. **36**, 301 (2008).
38) F. Di Benedetto, D. Borrini, A. Caneschi, G. Fornaciai, M. Innocenti, A. Lavacchi, C. A. Massa, G. Montegrossi, W. Oberhauser, L. a. Pardi, and M. Romanelli, Phys. Chemi. Miner. **38**, 483 (2011).
39) T. L. Evstigneeva, V. S. Rusakov, and Y. K. Kabalov, New Data Miner. **38**, 65 (2003).
40) W. Zalewski, R. Bacewicz, J. Antonowicz, A. Pietnoczka, T. L. Evstigneeva, and S. Schorr, J. Alloy. Compd. **492**, 35 (2010).
41) F. Izumi, K. Momma, Solid State Phenom. **130**, 15 (2007).
42) K. Momma, F. Izumi, J. Appl. Crystallogr. **44**, 1272 (2011).
43) H. J. Goldsmid, *Applications of Thermoelectricity* (Butler and Tanner, London, 1960).
44) P. Kubelka, F. Munk, Z. Tech. Phys. **12**, 593 (1931).
45) P. E. Blöchl, Phys. Rev. B **50**, 17953 (1994).
46) G. Kresse and D. Joubert, Phys. Rev. B **59**, 1758 (1999).
47) G. Kresse and J. Furthmüller, Comput. Mater. Sci. **6**, 15 (1996).
48) G. Kresse and J. Furthmüller, Phys. Rev. B **54**, 11169 (1996).
49) J. Heyd, G. E. Scuseria, and M. Ernzerhof, J. Chem. Phys. **118**, 8207 (2003).

50) H. J. Monkhorst and J. D. Pack, Phys. Rev. B **13**, 5188 (1976).

51) A. Suzumura, M. Watanabe, N. Nagasako, and R. Asahi, J. Electron. Mater. **43**, 2356 (2014).

52) Y. Liu, L. Zhao, Y. Liu, J. Lan, W. Xu, F. Li, B. Zhang, and D. Berardan, J. Am. Chem. Soc. **133**, 20112 (2011).

53) T.R. Wei, F. Li, and J.-F. Li, J. Electron. Mater. **43**, 2229 (2014).

54) Y. Goto, M. Tanaki, Y. Okusa, T. Shibuya, K. Yasuoka, M. Matoba, and Y. Kamihara, Appl. Phys. Lett. **105**, 022104 (2014).

55) C. D. Wagner, W. M. Riggs, L. E. Davis, J. F. Moulder, *Handbook of X-ray Photoelectron Spectroscopy* (Perkin-Elmer, Minnesota, 1979).

56) M. Scrocco, Chem Phys. Lett. **63**, 52 (1979).

57) J. J. Yeh, I. Lindau, At. Data Nucl. Data Tables **32**, 1 (1985).

58) E. J. Skoug, J. D. Cain, and D. T. Morelli, Appl. Phys. Lett. **98**, 261911 (2011).

59) V. I. Fitsul', *Heavily Doped Semiconductors* (Plenum Press, New York, 1969).

60) S. Chen, X. G. Gong, A. Walsh, and S. H. Wei, Appl. Phys. Lett. **94**, 041903 (2009).

61) H. Hiramatsu, I. Koizumi, K. B. Kim, H. Yanagi, T. Kamiya, M. Hirano, N. Matsunami, H. Hosono, J. Appl. Phys. **104**, 113723 (2008).

## Table Caption

Table 1. Refined structural parameters of $Cu_{3-x}SnS_4$ ($x_{nom}$ = 0.3).[a]

## Figure Captions

**Fig. 1.** (Color online) Crystallographic structure of CTS. Sites denoted by 2$b$ and 4$d$ are partially occupied by Cu and Sn.

**Fig. 2.** (Color online) XRD pattern and the Rietveld refinement of CTS. Crosses (red) and solid lines (black) represent the observed and calculated patterns, respectively. The blue lines denote the difference between observed and calculated patterns. The vertical marks (green) represent the Bragg diffraction angles of CTS, $Cu_{2-\delta}S$, and SnS from top to bottom, respectively. The black and red arrows indicate the diffraction due to $Cu_{2-\delta}S$ and SnS, respectively. The inset shows the SEM images.

**Fig. 3.** (Color online) Temperature dependence of electrical resistivity $\rho$ (top), Seebeck coefficient $S$ (middle), and thermal conductivity $\kappa$ (bottom) of CTS. The closed and open circles in $\rho$–T plot represent the heating and cooling data, respectively. The dashed line in $S$–$T$ plot is the guide to the eyes. The closed circles and open triangles in $\kappa$–$T$ plot denote the total thermal conductivity and lattice thermal conductivity, respectively.

**Fig. 4.** (Color online) Cu 2$p_{3/2}$ XPS spectrum of CTS, together with reported XPS spectra of reference materials.[52,53]

**Fig. 5.** (Color online) Valence-band XPS spectrum along with theoretical, total, and partial density of states (DOS) of CTS. Dashed vertical line denotes the Fermi energy. In the XPS spectra, the binding energy range of 0 to –0.4 eV cannot be observed with our equipment because of the calibration method used.

**Fig. 6.** (Color online) (a) Reflectivity and (b) ($\alpha h \nu s^{-1})^2$ spectra of CTS. Dotted line is a visual guide.

Table 1

| site | parameter | |
|---|---|---|
| 2a | $g_{Cu}{}^{b}$ | 1 |
|  | $U_{iso}$ ($10^{-5}$ nm$^2$) | 1.6(8) |
| 4d | $g_{Cu}{}^{c}$ | 0.699 |
|  | $g_{Sn}{}^{c}$ | 0.301(3) |
|  | $U_{iso}$ ($10^{-5}$ nm$^2$) | 2.6(3) |
| 2b | $g_{Cu}{}^{c}$ | 0.601 |
|  | $g_{Sn}{}^{c}$ | 0.399 |
|  | $U_{iso}$ ($10^{-5}$ nm$^2$) | 3.7(5) |
| 8i | $g_{S}{}^{b}$ | 1 |
|  | $x$ | 0.2465(6) |
|  | $z$ | 1.1241(10) |
|  | $U_{iso}$ ($10^{-5}$ nm$^2$) | 3.2(3) |

[a]Space group: $I\bar{4}2m$ (No. 121); Lattice parameters $a$ = 0.541752(8) nm and $c$ = 1.0784(3) nm, atomic position: 2a (0, 0, 0), 4d (0, 1/2, 1/4), 2b (0, 0, 1/2), 8i ($x$, $x$, $z$). The reliability factors and goodness of fit (GOF) were $R_{wp}$ = 4.017%, $R_B$ = 1.903%, and GOF = 1.2143.
[b]Site occupancies ($g$) for Cu on 2a site and S on 8i site were fixed at unity. [c]Constraints used for the $g$ of 4d and 2b sites: $g_{Cu}(4d) + g_{Sn}(4d) = 1$, $g_{Cu}(2b) + g_{Sn}(2b) = 1$, $2g_{Sn}(4d) + g_{Sn}(2b) = 1$.

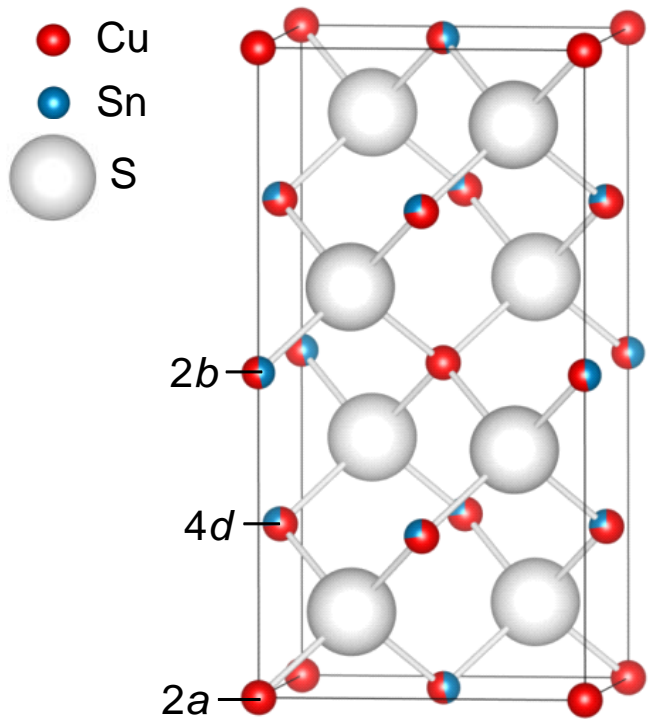

Fig.1. (Color Online)

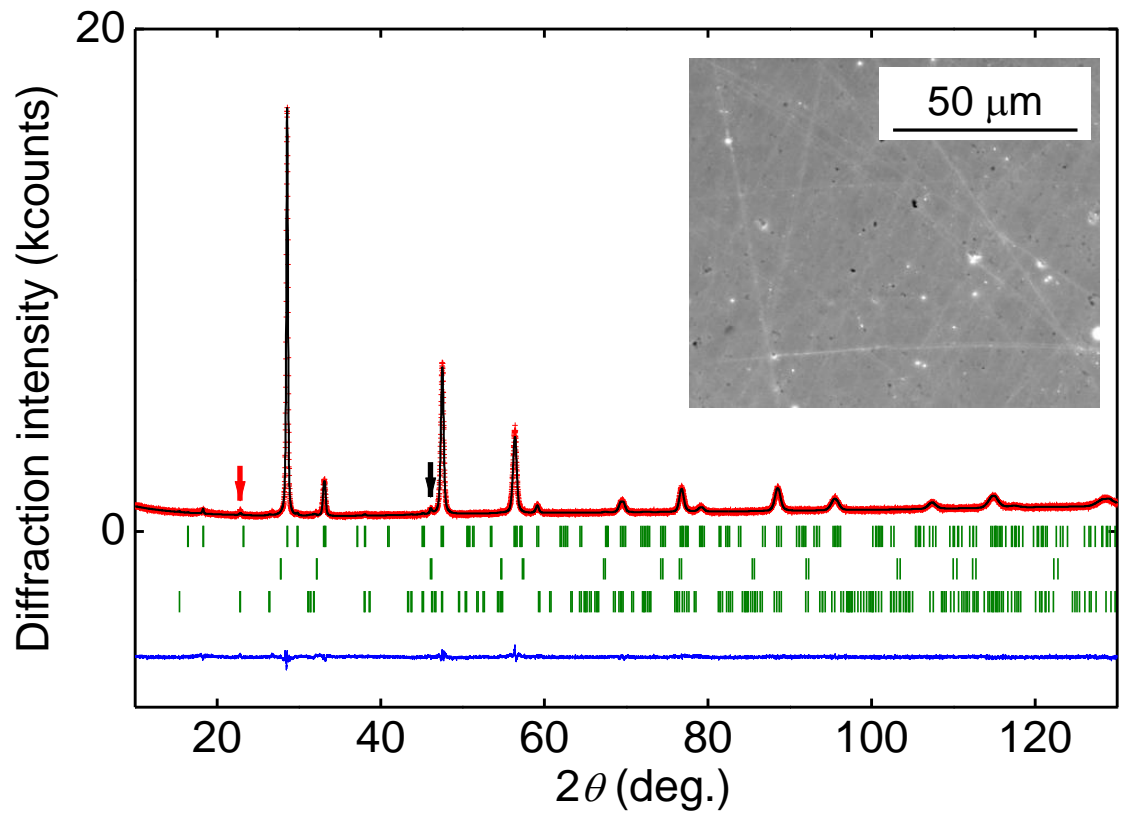

Fig. 2. (Color Online)

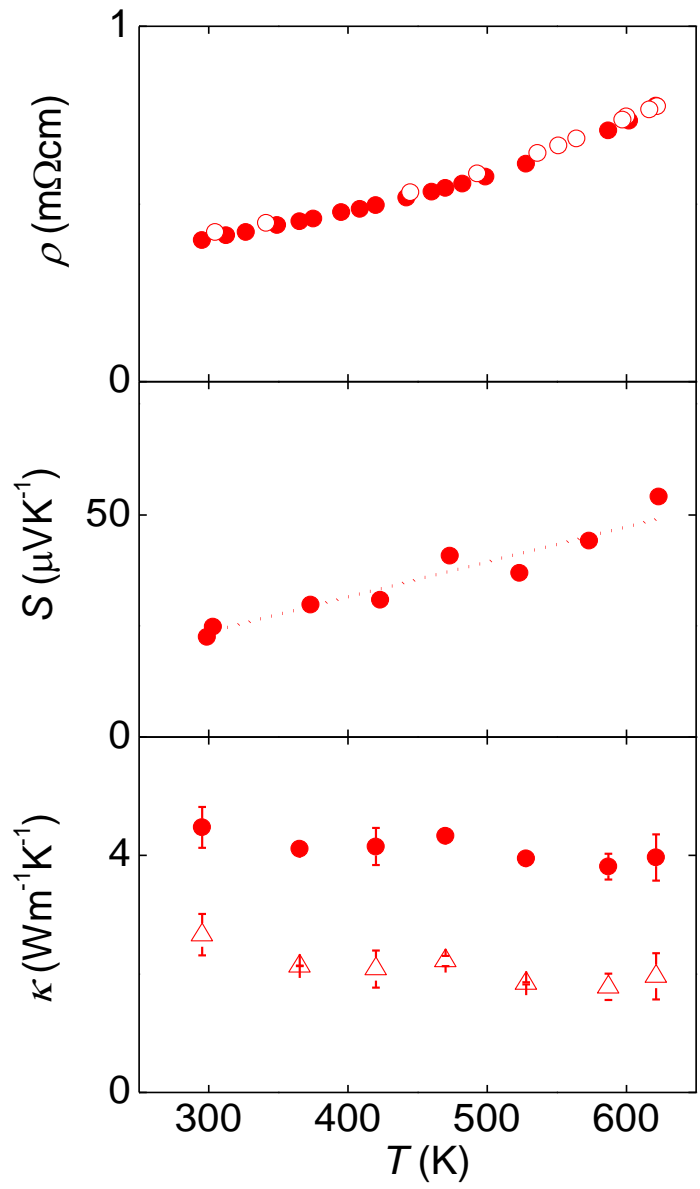

Fig. 3. (Color Online)

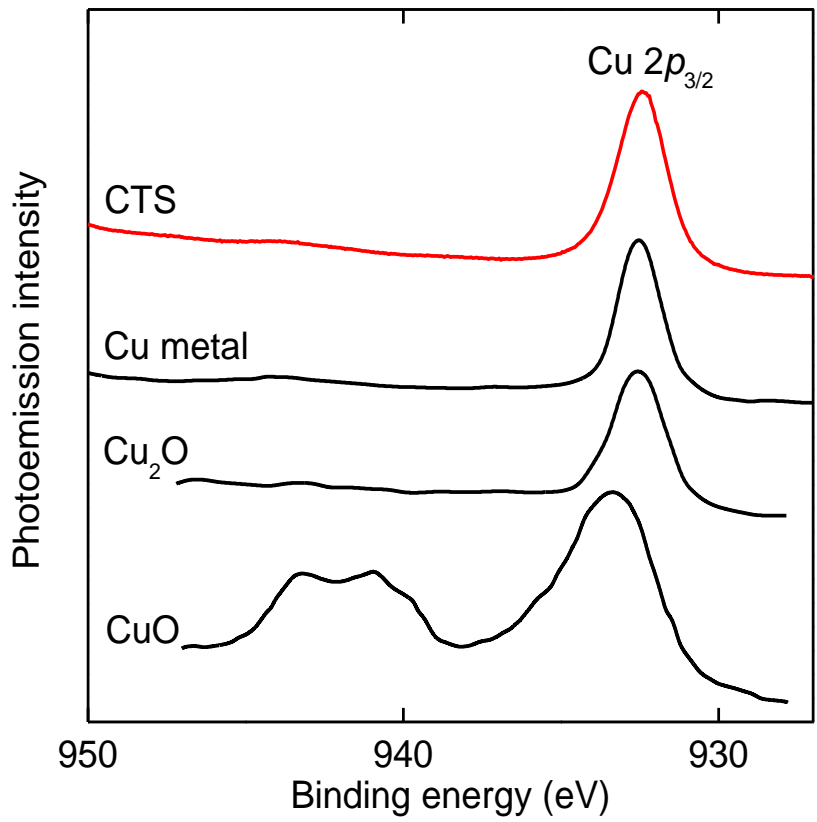

Fig. 4. (Color Online)

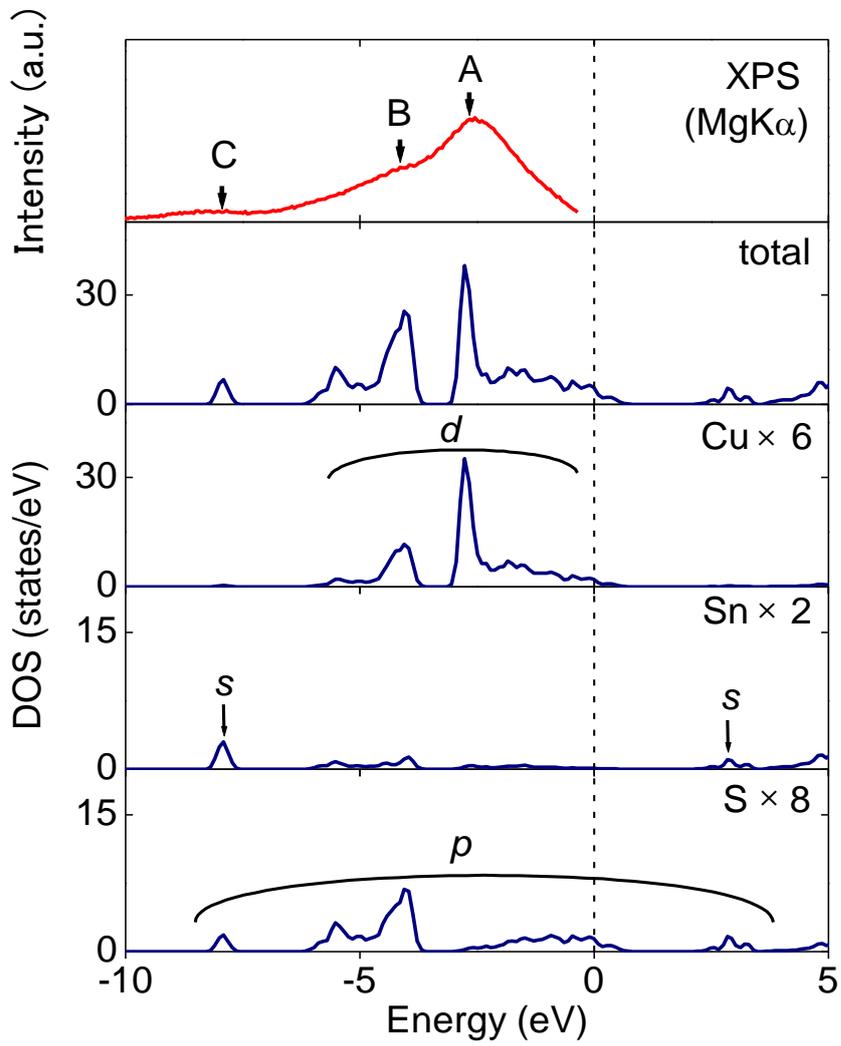

Fig. 5. (Color Online)

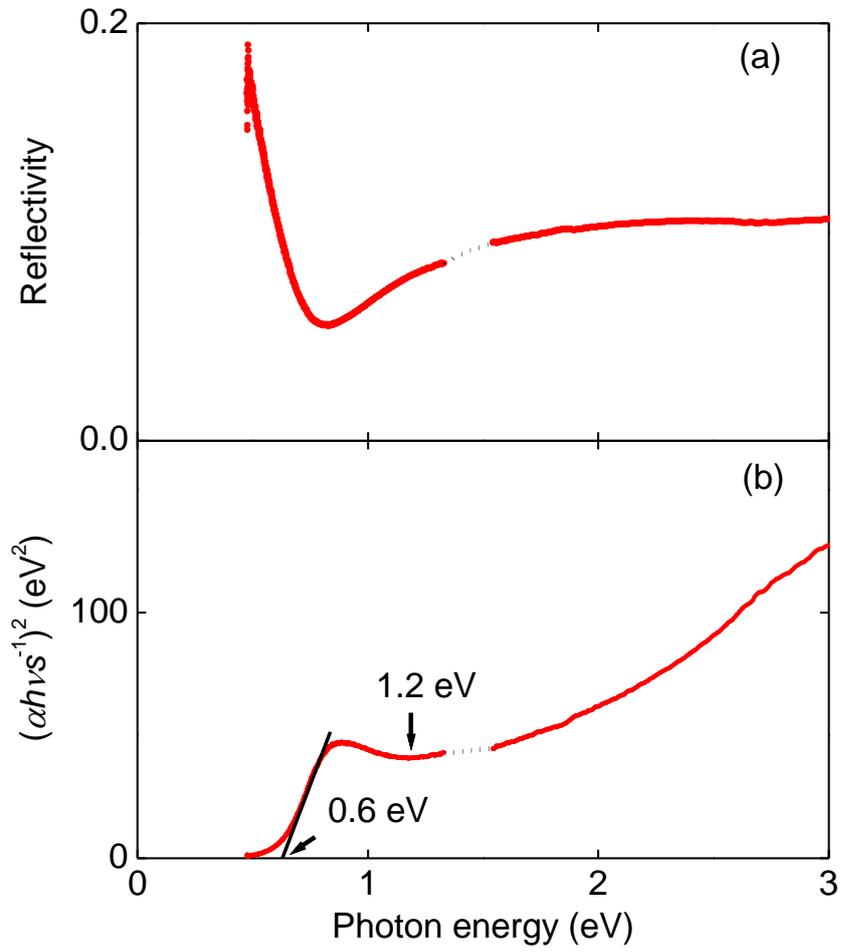

Fig. 6. (Color Online)